%% file: main.tex
\documentclass[conference]{IEEEtran}
\IEEEoverridecommandlockouts

\usepackage{cite}
\usepackage{amsmath,amssymb,amsfonts}
\usepackage{algorithmic}
\usepackage[dvipdfmx]{graphicx}
\usepackage{textcomp}
\usepackage{xcolor}
\usepackage{subfiles}
\usepackage{siunitx}
\usepackage{cleveref}
\usepackage{listings}
\usepackage{subcaption}
\usepackage{threeparttable}
\usepackage{multirow}
\usepackage{booktabs}
\usepackage{makecell}
\usepackage{physics}

\crefname{figure}{Fig.}{Fig.}
\crefname{table}{Table}{Table}
\crefname{section}{Sec.}{Sec.}
\crefname{lstlisting}{Listing}{Listing}

\def\BibTeX{{\rm B\kern-.05em{\sc i\kern-.025em b}\kern-.08em
    T\kern-.1667em\lower.7ex\hbox{E}\kern-.125emX}}
\begin{document}

\title{Dataflow-Oriented Classification and \\ Performance Analysis of GPU-Accelerated \\ Homomorphic Encryption}
\author{\IEEEauthorblockN{Ai Nozaki\IEEEauthorrefmark{1},
Takuya Kojima\IEEEauthorrefmark{2},
Hiroshi Nakamura\IEEEauthorrefmark{1}, and
Hideki Takase\IEEEauthorrefmark{1}
}

\IEEEauthorblockA{\IEEEauthorrefmark{1}\textit{The University of Tokyo}}
\IEEEauthorblockA{\IEEEauthorrefmark{2}\textit{Tsukuba University} }

\IEEEauthorblockA{\IEEEauthorrefmark{1}nozaki@hal.ipc.i.u-tokyo.ac.jp}
}

\maketitle

\begin{abstract}
Fully Homomorphic Encryption (FHE) enables secure computation over encrypted data, but its computational cost remains a major obstacle to practical deployment.
To mitigate this overhead, many studies have explored GPU acceleration for the CKKS scheme, which is widely used for approximate arithmetic.
In CKKS, CKKS parameters are configured for each workload by balancing multiplicative depth, security requirements, and performance.
These parameters significantly affect ciphertext size, thereby determining how the memory footprint fits within the GPU memory hierarchy.
Nevertheless, prior studies typically apply their proposed optimization methods uniformly, without considering differences in CKKS parameter configurations.
In this work, we demonstrate that the optimal GPU optimization strategy for CKKS depends on the CKKS parameter configuration.
We first classify prior optimizations by two aspects of dataflows which affect memory footprint and then conduct both qualitative and quantitative performance analyses.
Our analysis shows that even on the same GPU architecture, the optimal strategy varies with CKKS parameters with performance differences of up to 1.98 $\times$ between strategies, and that the criteria for selecting an appropriate strategy differ across GPU architectures.
\end{abstract}

\begin{IEEEkeywords}
Fully Homomorphic Encryption, CKKS, GPU
\end{IEEEkeywords}

\subfile{text/intro}

\subfile{text/background}

\subfile{text/proposal}

\subfile{text/evaluation}

\subfile{text/conclusion}

\bibliographystyle{IEEEtran}
\bibliography{reference}

\end{document}

%% file: text/intro.tex
\section{Introduction}

With the widespread use of cloud computing, protecting data confidentiality has become increasingly important.
FHE enables computation directly on encrypted data without decryption.
FHE supports arithmetic or logical operations over ciphertexts and is therefore expected to enable privacy-sensitive machine learning and data analytics.
However, the high computational overhead of FHE remains a significant barrier to its widespread adoption.
In the CKKS scheme~\cite{ckks17}, which is one of the most widely adopted HE schemes, execution times are $10^{4}$ to $10^{5}$ times longer than plaintext computations.
To mitigate this performance overhead, many prior studies have proposed GPU acceleration for CKKS.

In CKKS, parameters are configured for each workload by balancing the number of supported operations, security requirements, and performance.
We define the CKKS parameter set as a tuple $(dnum, N, L)$.
Here, $N$ denotes the polynomial degree of a ciphertext, and $L$ represents the maximum multiplicative depth.
Given $N$ and $L$, a CKKS ciphertext can be viewed as a structure consisting of $L$ chained polynomials of dimension $N$.
The parameter $dnum$ determines the number of partitions used in the KeySwitch operation, which accounts for approximately \SI{70}{\percent} of the total execution time in CKKS.
During KeySwitch, the $L$ polynomials are divided into $dnum$ parts, and each part is referred to as a digit.
These CKKS parameters are configured for each workload according to the aforementioned trade-offs.
For example, workloads requiring deep computational circuits, such as machine learning inference, use a larger $L$, which in turn necessitates a larger $N$ to satisfy security requirements.
Typical parameter ranges are $N = 2^{13}$ to $2^{17}$, $L = 0$ to $64$, and $dnum = 2$ to $10$.

Prior work on GPU acceleration of CKKS generally applies proposed optimization strategies uniformly, without considering differences in CKKS parameter configurations.
However, CKKS parameters significantly affect the memory footprint, which in turn impacts the effectiveness of optimizations.
Furthermore, under commonly used CKKS parameter ranges, the resulting memory footprint is close to the L2 cache capacity of modern GPUs.
This observation raises the question of whether a single optimization strategy can be effective across all CKKS parameter configurations.
For example, WarpDrive~\cite{warpdrive25} introduces a Parallelism-Enhanced Kernel that executes $dnum$ independent sub-operations in KeySwitch in parallel, effectively exploiting digit-level parallelism.
While this approach increases parallelism, it also enlarges the memory footprint, potentially causing frequent L2 cache evictions and performance degradation.
For a small parameter set, e.g., $(dnum, N, L) = (2, 2^{15}, 10)$, the on-chip memory footprint of WarpDrive's method is $2 \times 2^{15} \times 10 \times 8\ \mathrm{Bytes} = 5.12\ \mathrm{MB}$,
which fits within the L2 cache of modern GPUs (e.g., \SI{40}{\mega\byte} in NVIDIA A100).
In contrast, for a larger parameter set, e.g., $(4, 2^{16}, 50)$, the footprint reaches \SI{100}{\mega\byte}, exceeding the L2 cache capacity of many GPUs.
This observation suggests that the appropriate optimization strategy may depend on both CKKS parameters and the underlying GPU architecture.

Furthermore, prior work evaluates performance using only a limited number of CKKS parameter configurations.
For example, Cheddar~\cite{cheddar26} evaluates one configuration, WarpDrive~\cite{warpdrive25} evaluates five, and Neo~\cite{neo25} evaluates seven.
These represent only a small subset of the possible parameter combinations.
As a result, there is limited discussion on whether each optimization strategy remains effective across a broader range of CKKS parameters.

In this work, we demonstrate that the optimal GPU optimization strategy for CKKS varies according to the CKKS parameter configuration.
We classify existing optimization strategies into four categories and conduct both qualitative and quantitative performance analyses while varying CKKS parameters.

We first classify prior optimizations by two aspects of dataflow that affect memory footprint:
(1) whether digit-level parallelism is exploited in KeySwitch (DigitSerial or DigitParallel), and
(2) whether output granularity in KeySwitch is partitioned (OutputBulk or OutputChunked).
These two axes yield four distinct optimization strategies.
Based on the above classification, we observe that prior studies have primarily focused on optimizing the total DRAM access traffic, improving DRAM bandwidth utilization, and reducing the memory footprint.

To understand why the optimal optimization strategy changes with CKKS parameter configurations, we conduct a qualitative analysis of their computational characteristics.
We first estimate the on-chip memory footprint of each optimization strategy across CKKS parameter configurations.
Changes in memory footprint can lead to different cache behaviors, which in turn reshapes the distribution of execution stalls.
We employ GCoM~\cite{gcom22}, a GPU performance modeling framework, to analyze how these changes in memory footprint affect the overall performance of each optimization strategy.

Then, we perform quantitative performance evaluation across different CKKS parameter sets and GPU architectures.
Our results show that even on the same GPU architecture, the optimal optimization strategy varies depending on CKKS parameters, with performance differences up to 1.98$\times$ between strategies.
Moreover, the criteria for selecting the optimal strategy differ across GPU architectures.
Overall, we demonstrate that the appropriate optimization strategy depends on the combination of CKKS parameters and GPU architecture.
These findings suggest that CKKS parameter-aware optimization can unlock additional performance improvements for CKKS workloads accelerated on GPUs.

%% file: text/background.tex
\section{Background}
\label{sec:background}

\subsection{CKKS Encryption}
\label{subsec:ckks}

\subsubsection{CKKS}
CKKS~\cite{ckks17} is one of the FHE schemes that enable approximate addition and multiplication over encrypted real numbers.
Parameters used in CKKS are summarized in \cref{tab:ckksparam}.
In CKKS, a message is represented as a vector of $N/2$ real numbers.
Through the encode and encrypt procedures, the message is transformed into a ciphertext
$ct = (ct_0(x), ct_1(x)) \in R_Q^2$
, where $R_Q = \mathbb{Z}_Q[x]/(x^N - 1)$ denotes a degree-$N$ polynomial ring with coefficients in the integer ring $\mathbb{Z}_Q$, and $Q$ is the modulus.
Since $Q$ is a large integer (typically around \SI{1000}{\bit}), it is decomposed via the Chinese Remainder Theorem as $Q = \prod_{i=0}^{L-1} q_i$, where each $q_i$ is a machine-word-sized modulus.
As a result of this decomposition, a ciphertext has a structure consisting of $L$ chained polynomials, which is referred to as a ciphertext at level-$L$.
The level indicates the remaining multiplicative depth of the ciphertext, and it decreases by one after each multiplication.

The homomorphic operations in CKKS are defined as follows:
\begin{itemize}
  \item \texttt{HADD}: Adds two ciphertexts. $ct_{add}(ct, ct') = ct + ct'$
  \item \texttt{HMUL}: Multiplies two ciphertexts. $ct_{mul}(ct, ct') = (ct_0 ct'_0, ct_0 ct'_1 + ct'_0 ct_1) + \texttt{KS}(ct_1 ct'_1)$
  \item \texttt{HROT}: Rotates a ciphertext.
    $ct_{rot}(ct, r) = (0, \texttt{rot}(ct_1, r)) + \texttt{KS}(\texttt{rot}(ct_0, r))$
\end{itemize}

Here, \texttt{KS} denotes the KeySwitch operation, which transforms a ciphertext into a ciphertext under a different secret key.
In CKKS workloads, KeySwitch accounts for approximately \SI{70}{\percent} of the total execution time~\cite{pmlr-v162-lee22e}.
Accordingly, this work focuses on KeySwitch as the primary subject of analysis.

\subsubsection{KeySwitch}
\begin{figure}[tb]
    \centering
    \includegraphics[width=\linewidth]{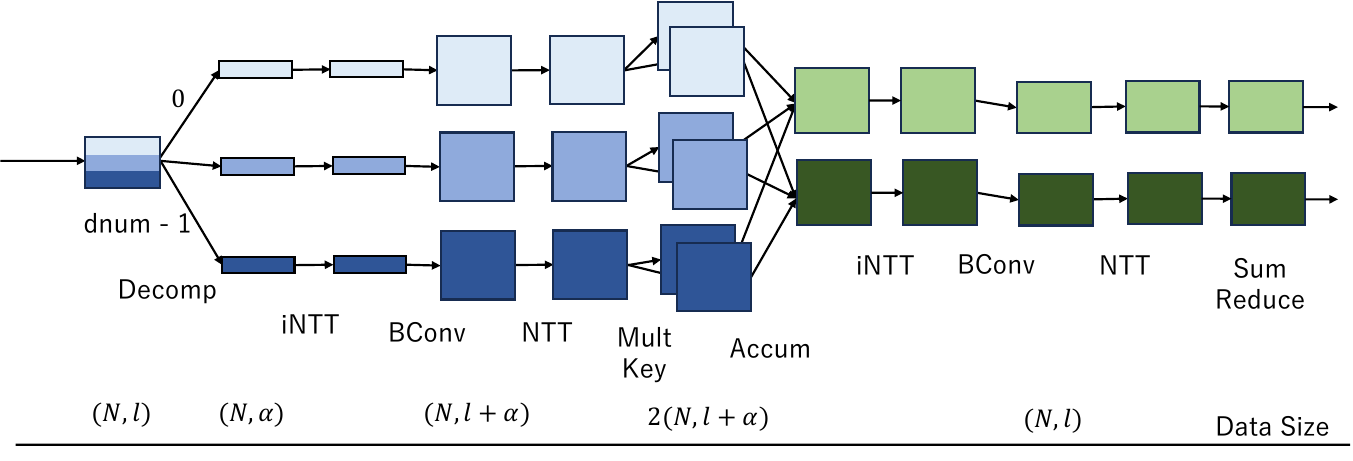}
    \caption{KeySwitch operation}
    \label{fig:keyswitch}
\end{figure}

The overall procedure of KeySwitch is illustrated in \cref{fig:keyswitch}. 
In KeySwitch, the input ciphertext is first decomposed into $dnum$ digits, each at level-$\alpha$, where $\alpha = (l + 1) / dnum$.
In the first phase, shown in blue, each digit is expanded to level $l + \alpha$ by performing the operations \texttt{iNTT} $\to$ \texttt{BConv} $\to$ \texttt{NTT}.
Each digit is then multiplied by the corresponding evaluation key, and the results are accumulated across all digits.
In the second phase, shown in green, the level of the ciphertext is reduced back to $l$ by again applying \texttt{iNTT} $\to$ \texttt{BConv} $\to$ \texttt{NTT}.
Here, \texttt{BConv} (Base Conversion) converts a polynomial into an approximately equivalent polynomial with a different set of moduli.
\texttt{NTT} (Number Theoretic Transform) is a variant of the Fast Fourier Transform defined over integer polynomial rings.
Ciphertexts are typically maintained in the NTT domain to accelerate polynomial multiplication.
However, since \texttt{BConv} requires coefficients in the standard (non-NTT) domain, \texttt{NTT} and its inverse (\texttt{iNTT}) are performed before and after \texttt{BConv}.
Because KeySwitch expands each digit to level $l + \alpha$, it increases both computational cost and data size. 

\subsubsection{CKKS Parameter Configuration}
These CKKS parameters are configured for each workload by balancing the supported multiplicative depth, security requirements, and performance.
A larger $Q$, i.e., a larger $L$, enables deeper computational circuits, however, it also weakens security.
To maintain a target security level, a larger $N$ is required when $L$ increases, which in turn raises the computational cost.
A smaller $dnum$ also reduces security, whereas a larger $dnum$ increases computational overhead and evaluation key size.
FHE compilers incorporate CKKS parameter selection into their compilation process~\cite{antace25,chet19}.
Similarly, FHE ASIC accelerators such as BTS~\cite{bts22} simulate execution time across a range of CKKS parameter configurations in advance to determine the supported parameter sets.

\begin{table}[tb]
    \centering
    \caption{Parameters used in CKKS}
    \label{tab:ckksparam}
    \begin{tabular}{cc} \toprule
        \textbf{Parameter} & \textbf{Description}\\ \midrule
        $N$ & Degree of polynomial  \\
        $Q$ & Modulus  \\
        $L$ & Maximum multiplicative level. $Q = \prod_{i=1}^{L} q_i$  \\
        $l$ & Current level $1 \le l \le L-1$ \\
        $dnum$ & Decomposition number \\
        $\alpha$ & $\lceil(L+1)/dnum\rceil$  \\
        \bottomrule
    \end{tabular}
\end{table}

\subsection{GPU Performance Modeling}
\label{subsec:gcom}

We briefly describe the GPU performance modeling method proposed in GCoM~\cite{gcom22}.
GCoM analytically estimates the execution cycles of a GPU kernel.

First, GCoM assumes a representative warp with no resource contention and calculates the execution cycles considering only stalls caused by data hazards.
The SASS (the instruction sequence of an NVIDIA GPU) of the representative warp is obtained via tracing.
Then, GCoM conducts interval analysis and divides the instruction sequence into intervals where instructions can be issued continuously and intervals where execution stalls due to data dependencies waiting for memory access or computational results.
Using cache simulation, the average memory access latency is estimated.
As a result of interval analysis, the stall cycles for each interval $S_k^{Intv}$ are derived.

Next, stalls caused by structural hazards among warps are considered to estimate the total execution cycles.
The total kernel execution cycles are computed as follows:
\begin{gather}
 C^{kernel} = \frac{\sum_{i=1}^{\#SM} C_i}{\#SM} \\
 C_i = \frac{\sum_{j=1}^{\#Subcore} C_{i,j}}{\#Subcore} + S_i
\end{gather}
where $C_{i,j}$ represents the execution cycles of subcore $j$ in SM $i$, and $S_i$ represents the stall cycles due to inter-subcore interference.
Here, a subcore refers to the execution unit associated with a GPU warp scheduler.

The execution cycles per subcore, $C_{i,j}$, are decomposed as:
\begin{align}
 C_{i,j} &= C_{i,j}^{Active} + C_{i,j}^{Idle} \\
         &= C_{i,j}^{Base} + S_{i,j}^{ComData} + S_{i,j}^{MemData} + C_{i,j}^{Idle}
\end{align}

$C_{i,j}^{Base}$ represents the ideal execution cycles assuming no stalls:
\begin{equation}
    C_{i,j}^{Base} = \frac{Warps \times InstsPerWarp}{IssueRate}
\label{eq:base_cycle}
\end{equation}

$S_{i,j}^{ComData}$ and $S_{i,j}^{MemData}$ represent stalls due to computational data dependencies and memory data dependencies, respectively.
They are calculated by subtracting the cycles hidden by other warps from the stall cycles $S_k^{Intv}$ for each interval as follows:
\begin{gather}
    S_{i,j}^{Com/MemData} = \sum_{k \in Intvs} \max(S_{k}^{Intv} - C_{other}, 0) \\
    C_{other} = P_{warp} \times (\#Warps - 1) \times AvgIntrvInsts
\label{eq:data_hazard_stall}
\end{gather}
where $P_{warp}$ denotes the probability that other warps can issue instructions during a stall interval caused by data hazards.

$C_{i,j}^{Idle}$ represents idle cycles incurred while waiting for other subcores.

The stall cycles due to inter-subcore interference $S_i$ are further decomposed as:
\begin{gather}
 S_i = S_i^{ComStruct} + S_i^{MemStruct} + S_i^{NoC} + S_i^{DRAM}
\end{gather}

$S_i^{ComStruct}$ represents stalls due to contention for computational units, and $S_i^{MemStruct}$ represents stalls due to L1 cache bank conflicts.
For each interval, the required cycles for each functional unit are calculated, and the difference between the maximum required cycles and the ideal issue cycles is accumulated as stall cycles.

$S_i^{NoC}$ and $S_i^{DRAM}$ represent stalls caused by memory access contention in the on-chip network (NoC) and DRAM, respectively.
They are estimated by multiplying the number of accesses (considering hit rates) by the respective access latencies:
\begin{gather}
  S_i^{NoC} = 0.5 \times \#SM \times M \times L^{NoC} \label{eq:noc_stall} \\
  S_i^{DRAM} = 0.5 \times \#SM \times M \times L2Miss \times L^{DRAM} \label{eq:dram_stall} \\
  M = (M^{Read} \times L1Miss + M^{Write}) \times \#Warps
\end{gather}

Here, $M$ denotes the number of memory accesses issued per warp.
$L^{DRAM}$ and $L^{NoC}$ denote the DRAM and NoC access latencies, respectively, which are calculated as:
\begin{gather}
  L^{DRAM} = f \times \frac{BlockSize}{Bandwidth^{DRAM}} \label{eq:dram_latency} \\
  L^{NoC} = f \times \frac{BlockSize}{Bandwidth^{NoC}}
\end{gather}

By computing these various stall cycles, the total kernel execution cycles and their breakdown can be estimated.

%% file: text/proposal.tex
\section{Qualitative Analysis}
In this study, we investigate the hypothesis that the appropriate optimization strategy for accelerating CKKS on GPUs depends on the chosen CKKS parameter configuration.
We first classify prior optimizations by two aspects of dataflows which affect memory footprint and then conduct qualitative performance analysis for each optimization strategy.


\subsection{Dataflow-Oriented Classification of CKKS Acceleration}
\label{subsec:existing_methods}

\begin{figure}[tb]
  \centering
  \begin{minipage}[b]{0.49\linewidth}
    \centering
    \includegraphics[width=\linewidth]{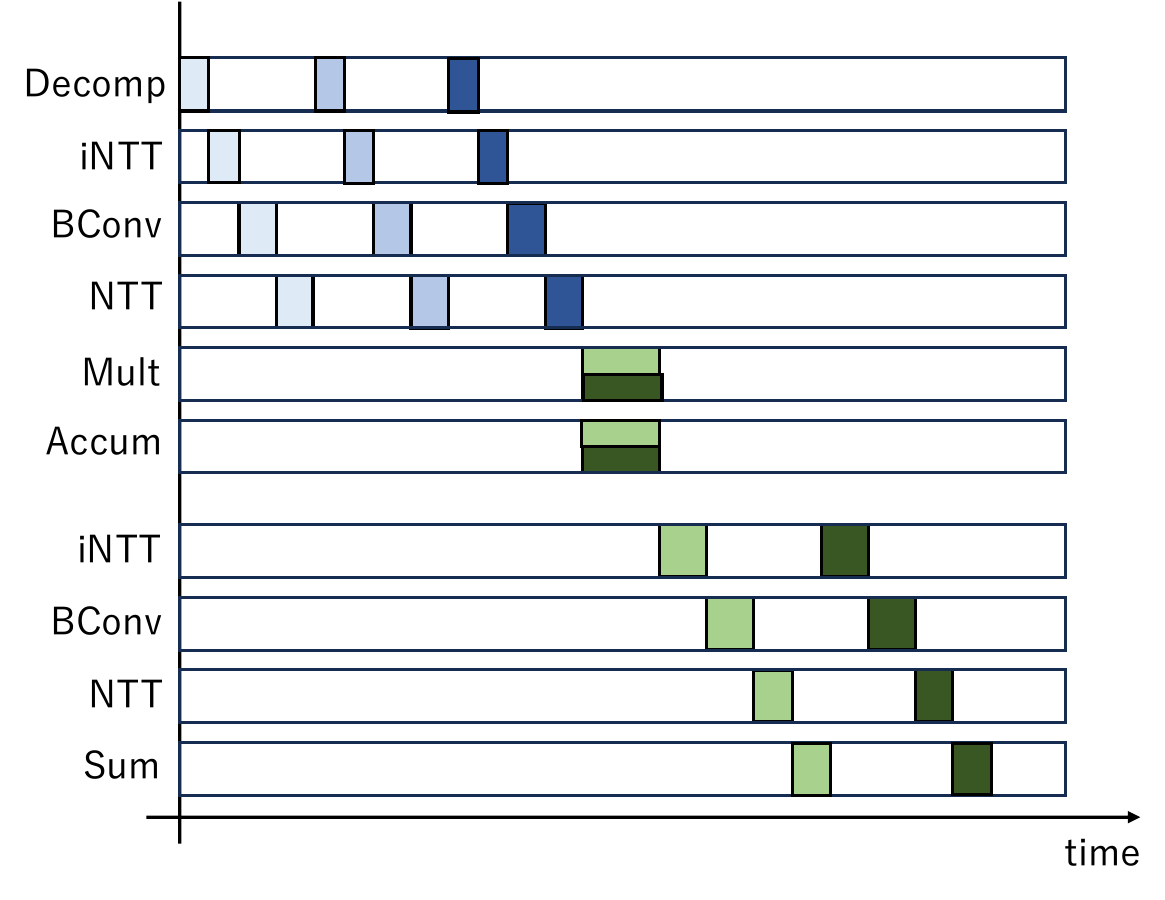}
    \subcaption{DSOB}
    \label{fig:dsob}
  \end{minipage}
  \hfill
  \begin{minipage}[b]{0.49\linewidth}
    \centering
    \includegraphics[width=\linewidth]{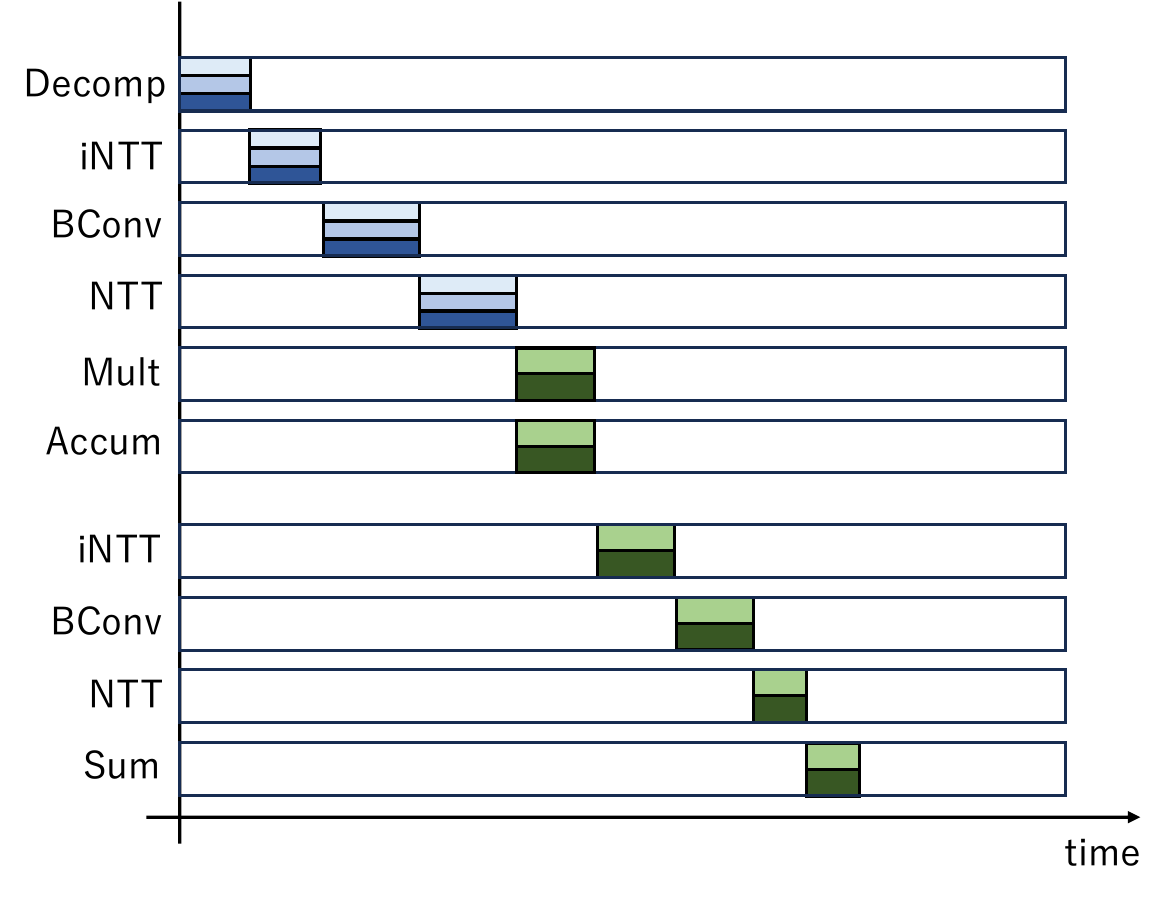}
    \subcaption{DPOB}
    \label{fig:dpob}
  \end{minipage}
  \hfill
  \begin{minipage}[b]{0.49\linewidth}
    \centering
    \includegraphics[width=\linewidth]{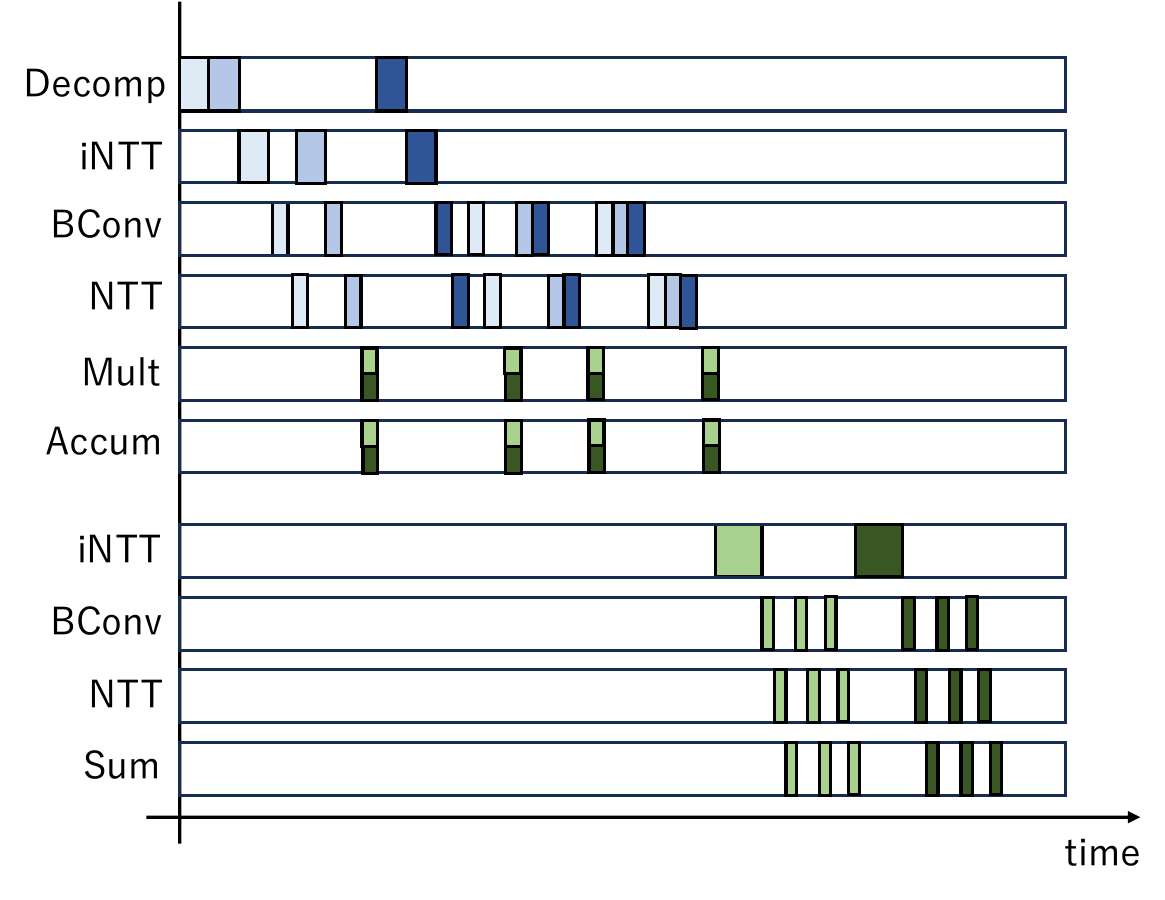}
    \subcaption{DSOC}
    \label{fig:dsoc}
  \end{minipage}
  \hfill
  \begin{minipage}[b]{0.49\linewidth}
    \centering
    \includegraphics[width=\linewidth]{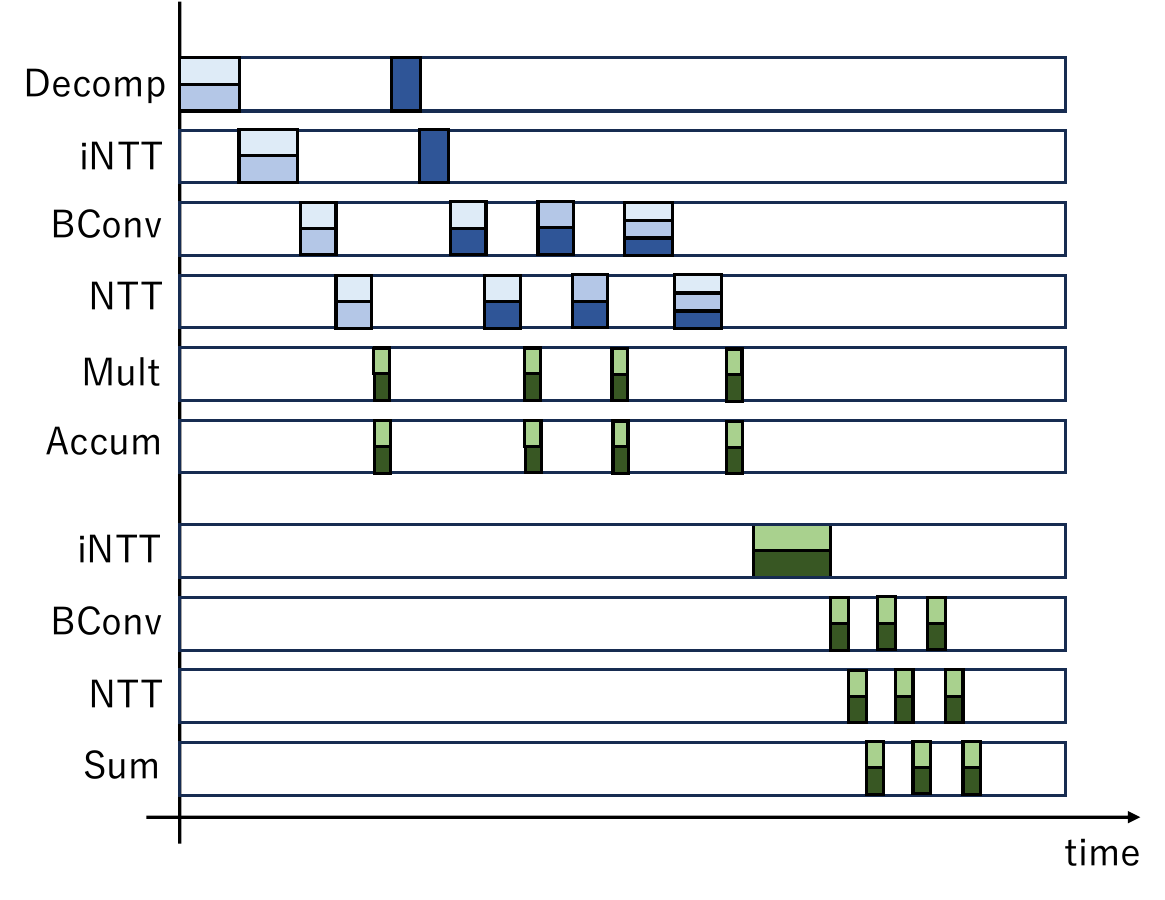}
    \subcaption{DPOC}
    \label{fig:dpoc}
  \end{minipage}
  \caption{KeySwitch dataflow in the four categories.}
  \label{fig:four_methods}
\end{figure}

We organize the dataflow-related optimization strategies proposed in prior studies on GPU acceleration.
Focusing on the dataflow of KeySwitch computation, we consider the following two axes:

\begin{enumerate}
  \item \textbf{Digit-level parallelism in KeySwitch.}
  KeySwitch consists of $dnum$ independent partial computations.
  We refer to the method that executes these computations sequentially as DigitSerial (DS), and the method that executes them in parallel within a single kernel as DigitParallel (DP).
  \item \textbf{Output granularity in KeySwitch.}
  KeySwitch produces a ciphertext of level-$(L+\alpha)$ and level-$L$ polynomials in the first and second phases, respectively.
  In the first and second phases, KeySwitch generates intermediate polynomials at levels $l+\alpha$ and $l$, respectively.
  We refer to the method that computes all levels at once as OutputBulk (OB), and the method that partitions the output and computes it incrementally as OutputChunked (OC).
  In OC, the number of output partitions is denoted by $chunks$.
\end{enumerate}

Combining these two axes yields four approaches.
\cref{fig:four_methods} illustrates the KeySwitch dataflow for these approaches.
In the figure, the horizontal axis represents time, indicating the execution order of the internal KeySwitch steps, and the colors correspond to those used in \cref{fig:keyswitch}.
For example, in DSOB, computations are performed sequentially across digits over time (DS), and for each digit, the entire output is computed in a single pass (OB).


\subsection{Mapping of Prior Work}
\label{subsec:ckks_gpu}

\begin{table}[tb]
  \centering
  \caption{Classification of prior work}
  \label{tab:existing_methods}
  \begin{tabular}{c|c|c} \toprule
    & DigitSerial & DigitParallel \\ \hline
    OutputBulk &
    \makecell[l]{%
      100x~\cite{100x21}, HE-Booster~\cite{he-booster-23}, \\
      FIDESlib~\cite{fideslib25}, Phantom~\cite{yang2024phantom}, \\
      CARM~\cite{carm23}, HEonGPU~\cite{heongpu24}, ~\cite{intelfhegpu22}
    } &
    \makecell[l]{%
      WarpDrive~\cite{warpdrive25},\\
      Cheddar~\cite{cheddar26}
    } \\ \hline
    OutputChunked & MAD~\cite{mad23} & - \\ \bottomrule
  \end{tabular}
\end{table}

This subsection maps prior work to the dataflow optimization taxonomy introduced above (\cref{tab:existing_methods}).
Since this work focuses on inter-kernel dataflow optimizations, we omit the details of intra-kernel computational optimizations (e.g., the use of Tensor Cores for NTT).

DSOB follows the baseline algorithmic dataflow without special optimizations.
100x~\cite{100x21} is the first work to accelerate the full CKKS pipeline on GPUs.
Identifying element-wise operations as a performance bottleneck, it introduced kernel fusion across adjacent kernels.
Since kernel fusion for element-wise operations has been adopted by nearly all subsequent studies, we exclude it from our classification and do not discuss it further.
HE-Booster~\cite{he-booster-23} is characterized by fine-grained inter-thread synchronization in NTT.
FIDESlib~\cite{fideslib25} provides an OpenFHE~\cite{OpenFHE22}-compatible interface.
Phantom~\cite{yang2024phantom} targets unified acceleration of BFV, BGV, and CKKS.
CARM~\cite{carm23} focuses on resource-constrained environments such as IoT.
~\cite{intelfhegpu22} presents optimizations tailored for Intel GPUs.

DPOB addresses the low utilization of computational resources in DSOB by extracting parallelism through kernel fusion across multiple digits in KeySwitch.
WarpDrive~\cite{warpdrive25} is the first work to introduce DPOB.
WarpDrive also proposes concurrent use of Tensor Cores and CUDA Cores for NTT.
Cheddar~\cite{cheddar26} makes a major contribution by introducing 25--30\,bit primes, and also proposes kernel fusion both inside and outside FHE operations from a dataflow perspective.

DSOC is proposed based on the observation that CKKS ASIC accelerators require substantial on-chip memory (e.g., \SI{512}{\mega\byte} as reported in BTS~\cite{bts22}).
MAD~\cite{mad23} partitions ciphertexts and reorders execution to maximize data reuse under limited on-chip memory capacity.
Although primarily motivated by ASIC design, MAD demonstrates its effectiveness via simulation on ASICs, GPUs, and FPGAs.
CiFlow~\cite{ciflow23} proposes a similar partitioned dataflow, but it is not GPU-focused.

DPOC is a theoretically feasible dataflow, but it has not yet been adopted in existing work.
This is likely because OC was originally proposed to reduce the required on-chip memory footprint for ASIC acceleration, whereas combining it with DP increases on-chip memory demand.
However, on GPU architectures where such an on-chip memory footprint is acceptable, DPOC may become effective as it conceptually lies between DPOB (high parallelism due to kernel fusion) and DSOC (high data reuse due to partitioning).
Therefore, we include DPOC in our analysis.

Several prior works do not fall into the four approaches described above.
TensorFHE~\cite{tensorfhe} is the first work to propose using Tensor Cores, and from a dataflow perspective it improves throughput by batching multiple FHE operations.
However, as also pointed out in WarpDrive~\cite{warpdrive25}, CKKS already processes $\frac{N}{2}$ real numbers in parallel within a single FHE operation, and thus workloads that can extract additional parallelism through batching are limited.
We therefore exclude batching from our classification.
Neo~\cite{neo25}, a successor to TensorFHE~\cite{tensorfhe}, proposes mapping not only NTT but also other kernels to matrix multiplications to leverage Tensor Cores.
Neo adopts the KLSS algorithm~\cite{ckksKeyDecompose23} for KeySwitch, but since this work targets the more widely used Hybrid algorithm~\cite{better20}, we do not include Neo in our classification.
GME~\cite{gme23} proposes architecture-level improvements based on AMD GPUs.
It supports NoC communication across Compute Units (CUs, analogous to NVIDIA Streaming Multiprocessors) and proposes scheduling that places blocks processing the same data on the same CU across kernels.
Because GME formulates scheduling as a general graph problem and does not clearly describe the resulting mapping, we exclude it from our classification.
Finally, HE-Booster~\cite{he-booster-23} also proposes multi-GPU execution strategies.
Since this work focuses on single-GPU optimizations, we exclude multi-GPU dataflow from our classification.


\subsection{Qualitative Performance Analysis}
\label{subsec:performance_modeling}

\begin{table}[tb]
  \centering
  \caption{Computational characteristics (denoting $dnum$ as $d$ and $chunks$ as $c$)}
  \label{tab:calculation_characteristics}
  \small
  \begin{tabular}{c|cccc} \toprule
     & DSOB & DPOB & DSOC & DPOC \\ \hline
    \makecell[c]{%
      on-chip memory\\
      footprint
    } & $\order{NL}$ & $\order{dNL}$ & $\order{\frac{NL}{c}}$ & $\order{\frac{dNL}{c}}$ \\
    kernel launches & $\order{d}$ & $\order{1}$ & $\order{dc}$ & $\order{c}$ \\
    warps/kernel & $\order{1}$ & $\order{d}$ & $\order{\frac{1}{c}}$ & $\order{\frac{d}{c}}$ \\
    instructions/warp & $\order{1}$ & $\order{1}$ & $\order{1}$ & $\order{1}$ \\
    \bottomrule
  \end{tabular}
\end{table}

\begin{figure}[tb]
    \centering
    \includegraphics[width=\linewidth]{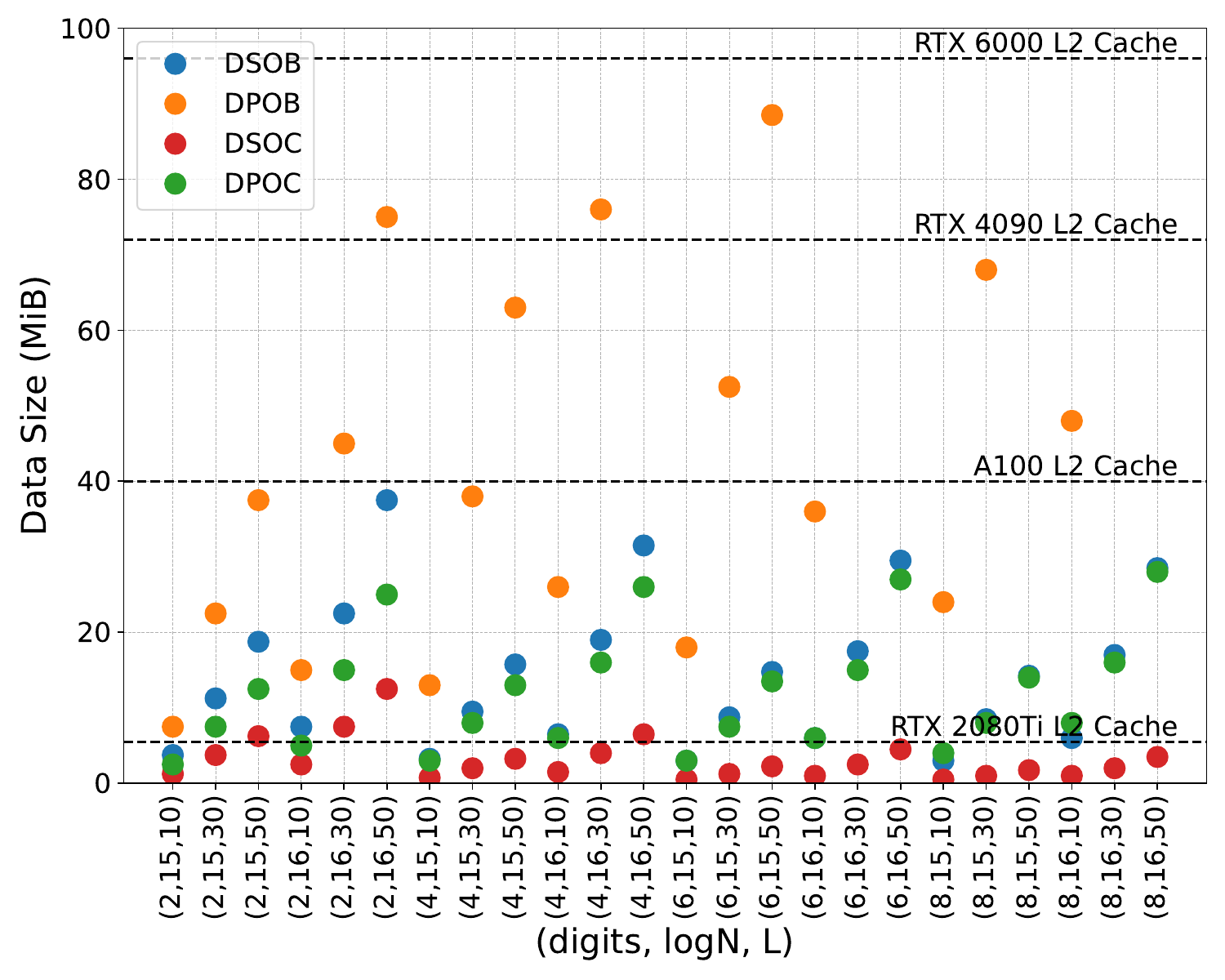}
    \caption{Relationship between memory footprint and CKKS parameter}
    \label{fig:memory_usage}
\end{figure}

We qualitatively analyze the performance of the classified approaches using the GPU performance modeling framework introduced in \cref{subsec:gcom}.

\cref{tab:calculation_characteristics} summarizes the differences in computational characteristics among the four approaches.
First, memory footprint and cache behavior vary across approaches.
\cref{fig:memory_usage} compares the memory footprint determined by the CKKS parameters with the L2 cache capacities of several GPUs. 
Since a ciphertext consists of $L$ polynomials of dimension $N$, its size scales as $\mathcal{O}(NL)$. 
In DP, the footprint increases by a factor of $dnum$ because it concurrently processes $dnum$ ciphertext components that are handled separately in DS. 
In OC, the footprint decreases by a factor of $1/chunks$ because the ciphertext is processed in $chunks$ partitions.
The DSOC footprint (shown in red) fits within the L2 cache for many CKKS parameter settings, even on GPUs with relatively small L2 caches such as the RTX~2080~Ti with \SI{4}{\mega\byte}.
In contrast, the DPOB footprint (shown in orange) can reach several tens of \si{\mega\byte}, and for some CKKS parameter settings it exceeds the L2 cache capacity even on high-end GPUs.
Consequently, the cache hit rate varies depending on the relationship between the memory footprint and the L2 cache capacity.

The four approaches also differ in kernel granularity and the number of kernel launches.
In DP, the $dnum$ kernels used in DS are merged into a single kernel. 
Therefore, the number of kernel launches is reduced to $1/dnum$ times of that in DS.
In contrast, OC divides a single kernel in OB into $chunks$ number of smaller kernels, increasing the number of kernel launches by a factor of $chunks$.
Because the total amount of computation remains identical across the four approaches, the computation per kernel scales inversely with the number of launches.
Although the mapping of computations to warps within each kernel is implementation-dependent, we assume that the number of executed instructions per warp remains constant and that the number of warps is adjusted accordingly.
That is, DP increases the number of warps per kernel by a factor of $dnum$ compared with DS, while OB increases the number of warps per kernel by a factor of $chunks$ compared with OC, such that the product of the number of launches and the number of warps per launch remains constant.

These differences in computational characteristics lead to the following performance implications.
\begin{itemize}
  \item Ideal instruction execution cycles \mbox{}\\
  The ideal execution cycles without stalls, $C_{i,j}^{Base}$, are identical across the four approaches.
  This is because $InstsPerWarp$ in \cref{eq:base_cycle} is constant, and the total number of warps executed over time is the same for all approaches.

  \item Stalls due to data hazards \mbox{}\\
  Stalls caused by data hazards, $S^{ComData}$ and $S^{MemData}$, depend on the cache hit rate and the number of concurrently executing warps.
  In \cref{eq:data_hazard_stall}, a higher cache hit rate reduces the stall cycles $S_k^{Intv}$ incurred while waiting for memory accesses.
  Furthermore, a larger number of concurrent warps enables more effective latency hiding, as data-hazard stalls in one warp can be overlapped with the execution of other warps.
  
  DP and OB exhibit a high warp density as long as the number of simultaneously resident blocks per SM remains within the architectural limit, resulting in better latency hiding.
  DP tends to require large memory footprint and may lower the cache hit rate, leading to increased stalls.
  Conversely, when the footprint fits within the L2 cache, DPOB can benefit from both high cache hit rates and effective latency hiding due to its high warp density.

  \item Stalls due to structural hazards \mbox{}\\
  Structural-hazard stalls caused by inter-subcore interference, $S^{ComStruct}$ and $S^{MemStruct}$, increase with the number of concurrently executing warps.
  Consequently, DP incurs more such stalls than DS, and OB incurs more than OC.

  \item Stalls due to NoC/DRAM access contention \mbox{}\\
  As shown in \cref{eq:noc_stall} and \cref{eq:dram_stall}, $S^{NoC}$ and $S^{DRAM}$ increase with the number of concurrently executing warps.
  Therefore, similar to structural hazards, these stalls are larger in DP than in DS and larger in OB than in OC.
  In addition, these stalls increase when the L1/L2 cache hit rates are low.
  Hence, for approaches whose memory footprint exceeds the L2 cache capacity (as illustrated in \cref{fig:memory_usage}), $S^{NoC}$ and $S^{DRAM}$ become more significant.
  We expect the impact of the L1 cache hit rate to be limited, because differences in inter-kernel dataflow do not substantially alter L1 cache utilization within each kernel.
  
  \item Kernel launch overhead \mbox{}\\
  Since kernel launch overhead increases with the number of kernel launches, the overhead is larger in DSOC than in DSOB, and in DPOC than in DPOB.
\end{itemize}

In summary, the four approaches exhibit distinct computational characteristics, each of which may either increase or decrease execution time depending on the type of stall involved.

%% file: text/evaluation.tex
\section{Quantitative Analysis}
\label{sec:evaluation}

\begin{table}[tb]
  \centering
  \caption{GPUs used in the evaluation}
  \label{tab:evaluation_environment}
  \begin{tabular}{c|cccc} \toprule
    GPU & Peak INT32 & L2 cache & Frequency 
    & \makecell[c]{%
      DRAM \\
      bandwidth
    } \\ \hline
    RTX 6000 Ada & \SI{44.5}{TOPS} & \SI{96}{\mega\byte} & \SI{2.51}{GHz} & \SI{960}{GB/s} \\ 
    RTX 4090 & \SI{41.3}{TOPS} & \SI{72}{\mega\byte} & \SI{2.52}{GHz} & \SI{1008}{GB/s} \\
    A100 & \SI{19.5}{TOPS} & \SI{40}{\mega\byte} & \SI{1.41}{GHz} & \SI{1555}{GB/s} \\ 
    RTX 2080 Ti & \SI{13.4}{TOPS} & \SI{5.5}{\mega\byte} & \SI{1.67}{GHz} & \SI{616}{GB/s} \\
    \bottomrule
  \end{tabular}
\end{table}

\subsection{Methodology}

We next conduct a quantitative performance analysis of the four approaches.
For the CKKS parameter configurations, we consider
$N \in \{2^{14}, 2^{15}, 2^{16}, 2^{17}\}$,
$L \in \{10, 30, 50\}$, and
$dnum \in \{2, 4, 6, 8\}$.
Note that the parameter combination $(L, dnum) = (10, 8)$ does not meet the security requirements of CKKS and is therefore excluded from the evaluation.

The GPUs used in our experiments are summarized in \cref{tab:evaluation_environment}.
We evaluate four GPUs representing different classes of computing resources.
The RTX 6000 Ada server is equipped with an Intel Xeon Gold 6226R CPU and \SI{155}{\giga\byte} of memory.
The RTX 4090 server features an Intel Xeon w9-3495X CPU and \SI{502}{\giga\byte} of memory.
The A100 server is equipped with an Intel Xeon Gold 6226R CPU and \SI{187}{\giga\byte} of memory.
The GTX 2080 Ti server features an Intel Xeon Gold 6240R CPU and \SI{251}{\giga\byte} of memory.
For profiling, we use NVIDIA Nsight Compute~\cite{nsight}.

Under these experimental conditions, we measure the execution time of homomorphic multiplication (\texttt{HMUL}).
Each experiment is repeated 100 times, and we report the average execution time.
The reported time includes the entire \texttt{HMUL} process, including data transfer from the CPU to the GPU.

\begin{figure}[tb]
    \centering
    \includegraphics[width=0.9\linewidth]{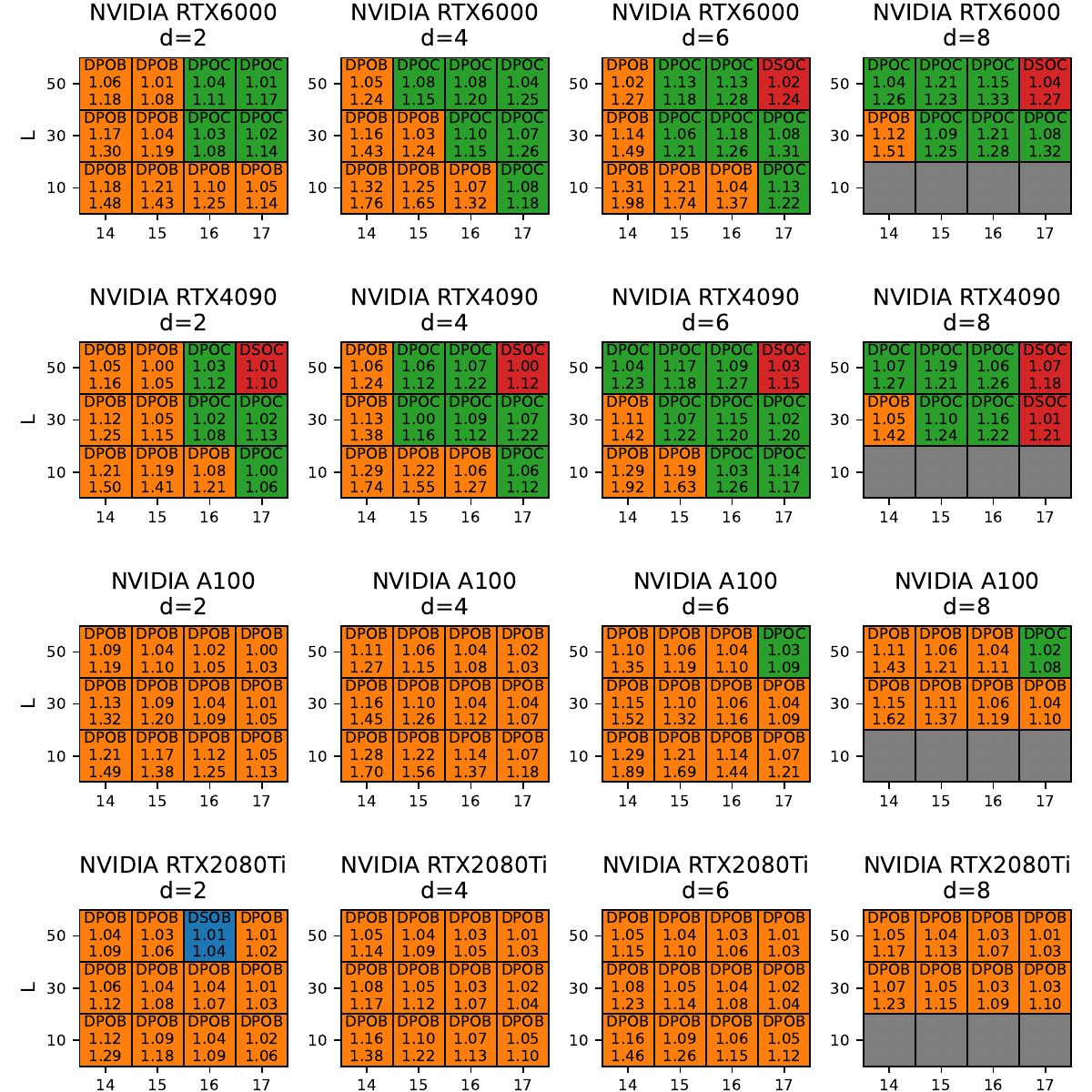}
    \caption{Distribution of the approach that achieves the best execution time. For each cell, the top row indicates the best-performing approach, and the middle/bottom rows show the performance ratio relative to the second-best/worst-performing strategies.}
    \label{fig:best_dataflow}
\end{figure}

\begin{figure*}[tb]
    \centering
    \includegraphics[width=\linewidth]{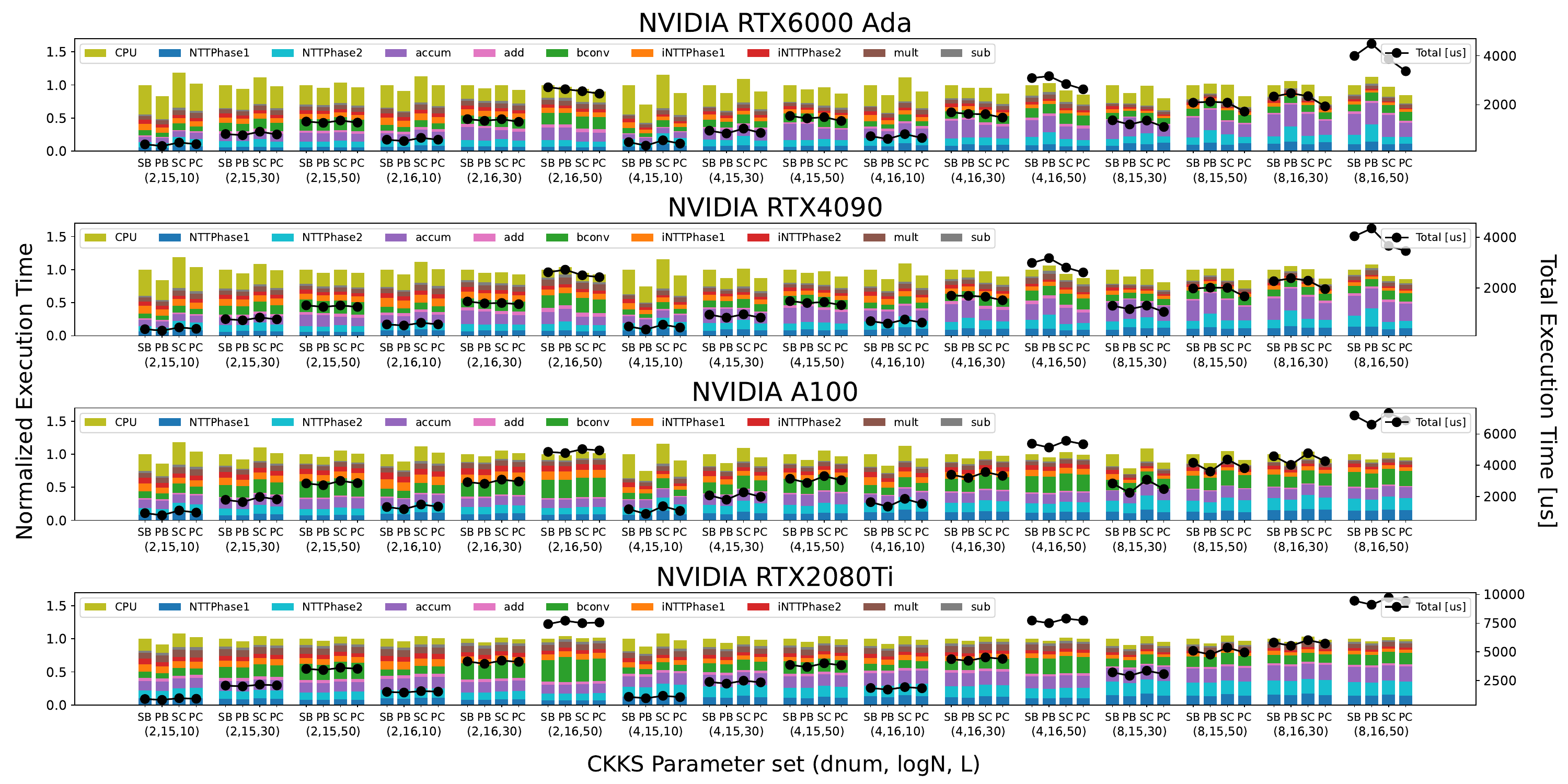}
    \caption{Execution times across optimization strategies. S/P denote DigitSerial/Parallel, and B/C denote OutputBulk/Chunked.
    The stacked bars show the breakdown of execution time, normalized to DSOB.
    The line plots indicate the absolute execution time.}
    \label{fig:execution_time}
\end{figure*}

\begin{figure}[htb]
  \centering
  \begin{minipage}[b]{\linewidth}
    \centering
    \includegraphics[width=\linewidth]{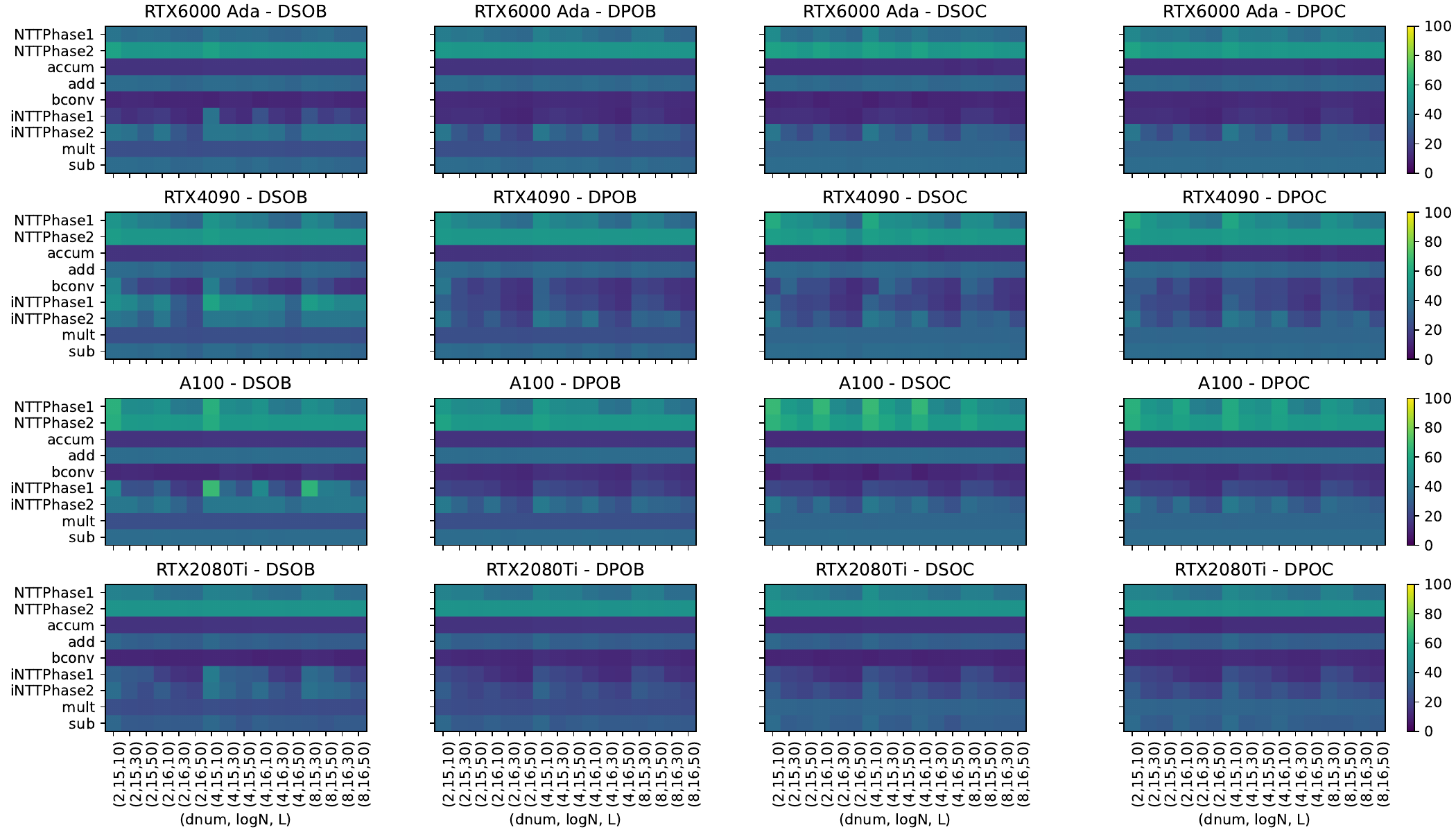}
    \subcaption{L1 cache}
    \label{fig:l1_cache_hit}
  \end{minipage}
  \hfill
  \begin{minipage}[b]{\linewidth}
    \centering
    \includegraphics[width=\linewidth]{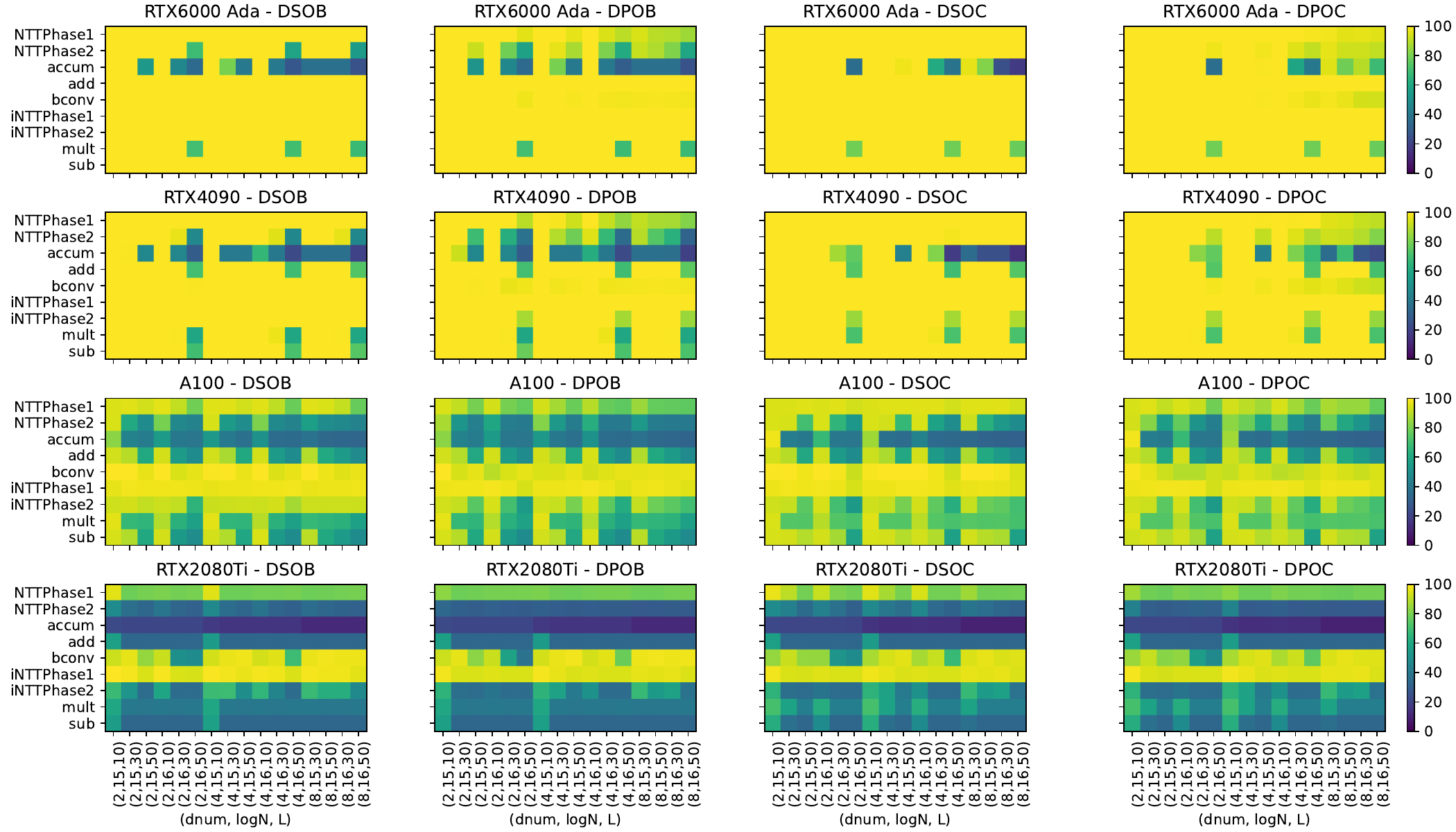}
    \subcaption{L2 cache}
    \label{fig:l2_cache_hit}
  \end{minipage}
  \caption{Cache hit rates}
  \label{fig:cache_hit}
\end{figure}

\begin{figure}[htb]
  \centering
  \begin{minipage}[b]{0.49\linewidth}
    \centering
    \includegraphics[width=\linewidth]{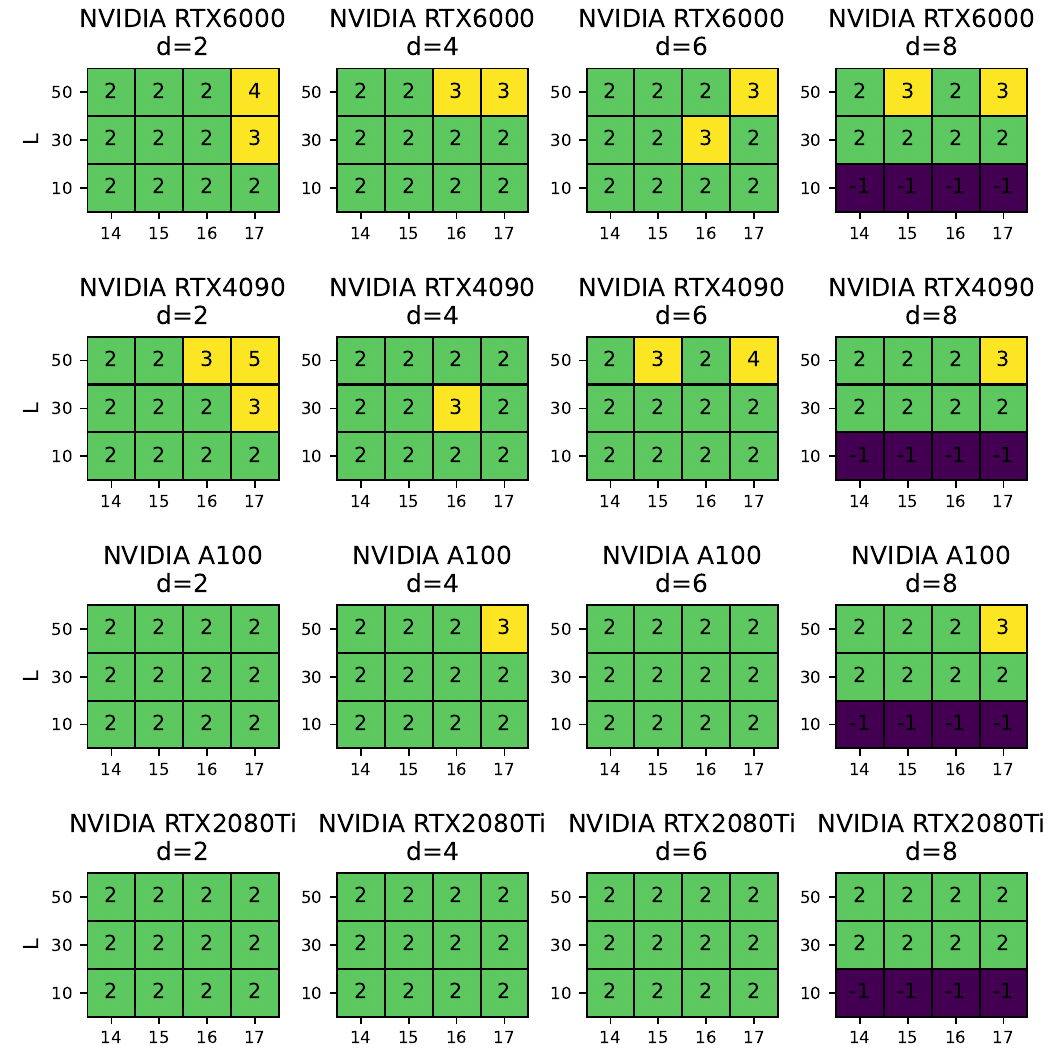}
    \subcaption{DSOC}
    \label{fig:best_chunks_dsoc}
  \end{minipage}
  \hfill
  \begin{minipage}[b]{0.49\linewidth}
    \centering
    \includegraphics[width=\linewidth]{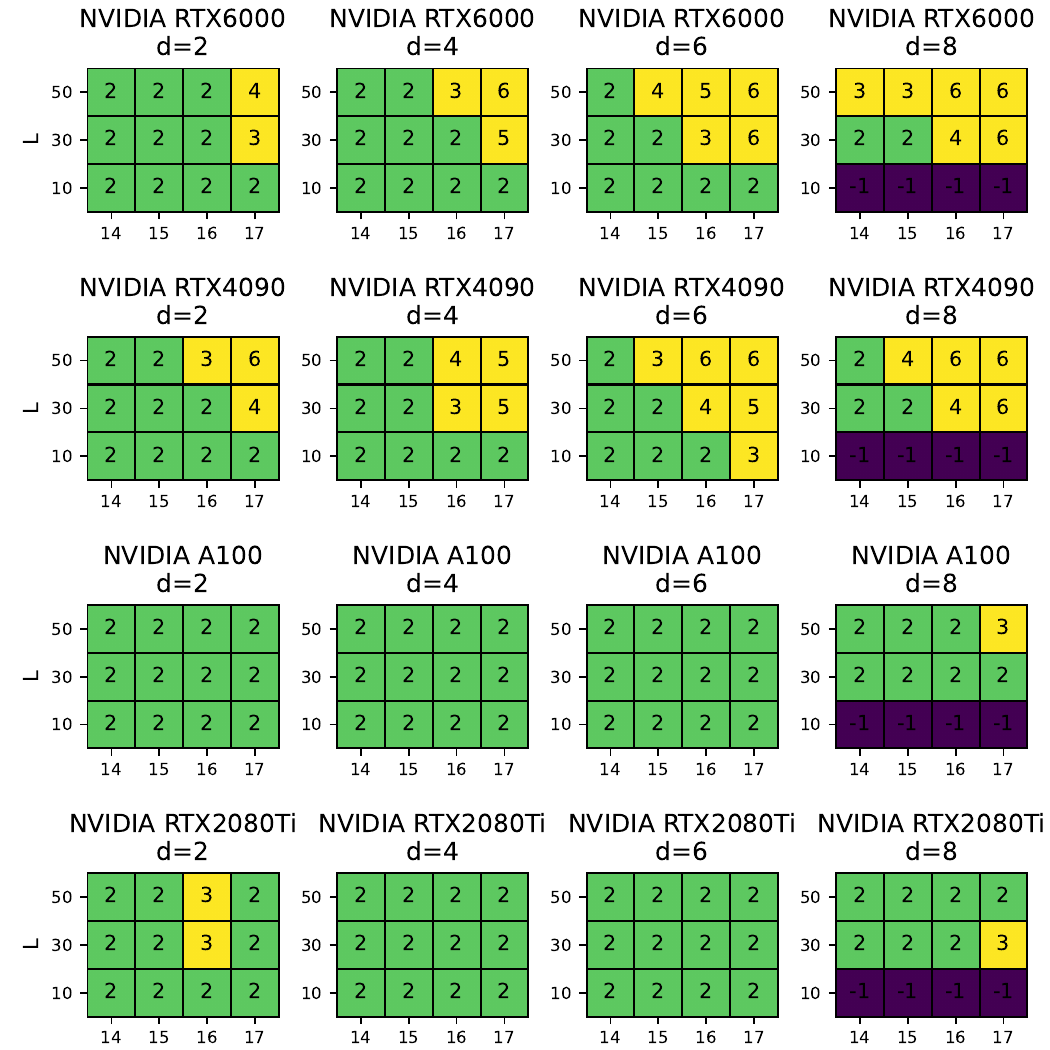}
    \subcaption{DPOC}
    \label{fig:best_chunks_dpoc}
  \end{minipage}
  \caption{Best chunks for OutputChunked}
  \label{fig:best_chunks}
\end{figure}

\begin{figure}[tb]
    \centering
    \includegraphics[width=\linewidth]{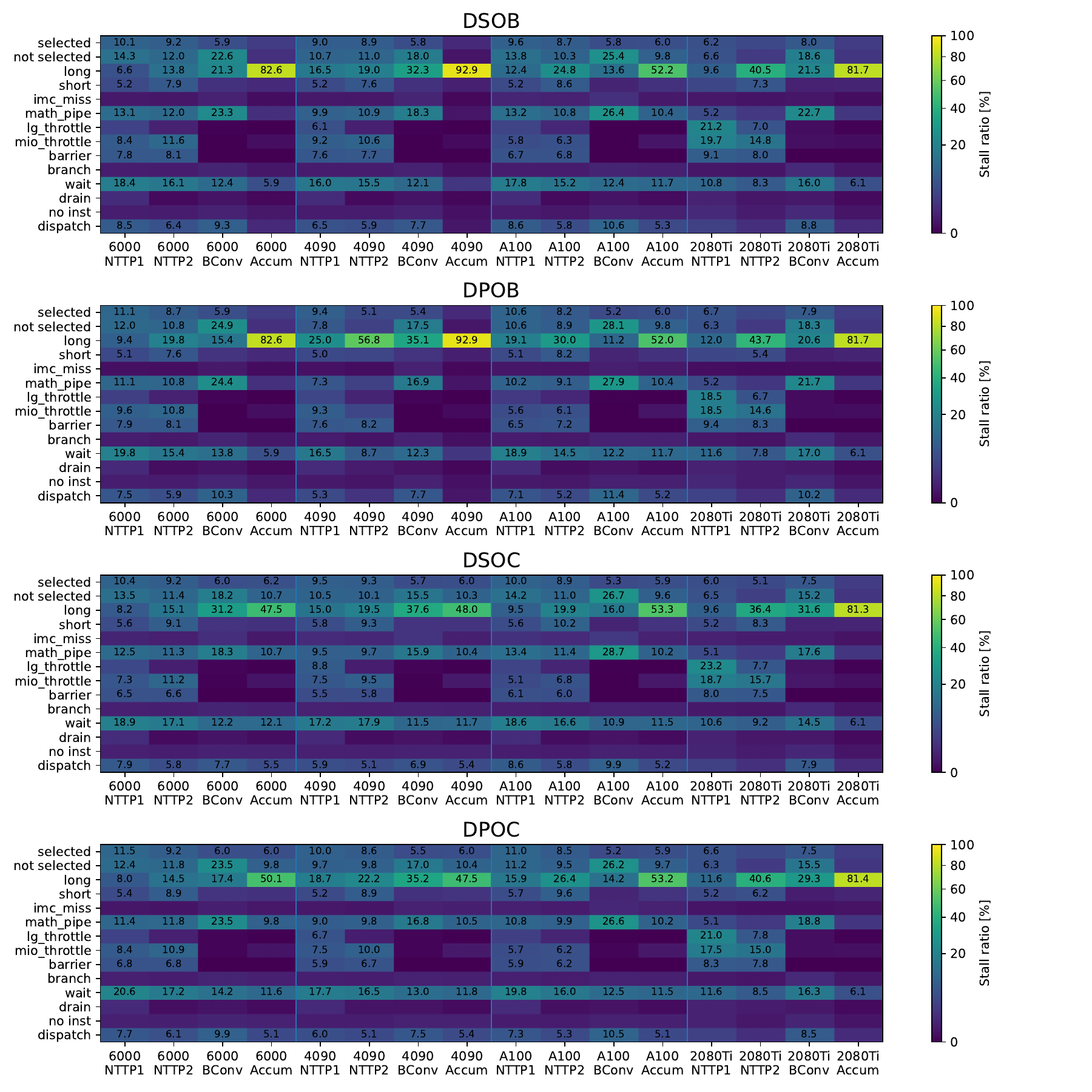}
    \caption{Breakdown of stall cycles with $(dnum, N, L) = (4, 2^{16}, 30)$}
    \label{fig:stall_breakdown}
\end{figure}

\subsection{Execution Time across CKKS Parameters}
\cref{fig:best_dataflow} shows the distribution of optimization strategies that achieve the best execution time.
\cref{fig:execution_time} presents the detailed execution times and their breakdown.
The stacked bars represent execution time normalized to DSOB, while the line plots indicate absolute execution time.

On the RTX 6000 Ada and RTX 4090, the optimal strategy varies with the CKKS parameters even on the same GPU.
Specifically, as the CKKS parameters increase, the best-performing strategy shifts from DPOB (orange) to DPOC (green) and then to DSOC (red).
This trend corresponds to the ordering of the on-chip memory footprints of the approaches. 
When the L2 cache capacity becomes less than about twice the footprint, the optimal strategy tends to shift to the approach with the next smaller footprint.
The maximum performance gap between the best and worst strategies reaches 1.98$\times$ on the RTX 4090 at $(dnum, N, L) = (6, 2^{14}, 10)$.
For small parameter sets (e.g., $(2, 2^{15}, 30)$ and $(4, 2^{15}, 10)$), DPOB achieves the best performance, while DSOC exhibits significantly longer \texttt{NTT} and \texttt{BConv} execution times than the other strategies.
Furthermore, kernel launch overhead (labeled as \texttt{CPU}) accounts for a substantial fraction of the total time relative to GPU computation, making DSOC particularly costly in these cases.
In contrast, for larger parameter sets, DPOB becomes the worst-performing strategy.
For example, at $(4, 2^{16}, 50)$ and $(8, 2^{16}, 50)$, the execution time of DPOB increases substantially, particularly in \texttt{NTTPhase2}, and DPOC achieves the best performance.

On the A100, DPOB achieves the best execution time for most CKKS parameter settings.
As shown in \cref{fig:execution_time}, the relative ordering of execution times among the four strategies is generally DPOB, DPOC $<$ DSOB, DSOC across most parameter sets.
Regardless of parameter size, DSOC tends to incur longer kernel execution and launch times than the other strategies.
In a few cases, DPOC performs best, but the difference from the worst strategy is less than 10\%, indicating only a marginal gap.

On the RTX 2080 Ti, DSOB provides the best performance for most CKKS parameters.
The performance differences among the four strategies are smaller than on the A100, and aside from kernel launch overhead, execution times are nearly identical.

Overall, DPOB achieves the best execution time under a wide range of conditions on the A100 and RTX 2080 Ti.
On the A100, we attribute this to architectural features that mitigate DPOB’s primary drawback, which is stalls caused by DRAM access contention ($S^{DRAM}$).
Because DPOB has a large on-chip footprint, it is more likely to incur DRAM accesses as a result of L2 cache misses.
Indeed, \cref{fig:l2_cache_hit} shows that even on the A100, DPOB tends to have a lower L2 cache hit rate than the other strategies.
As shown in \cref{eq:dram_stall}, $S^{DRAM}$ increases proportionally to the number of L2 cache misses, making it more pronounced for DPOB.
However, $S^{DRAM}$ is also proportional to DRAM access latency,
$L^{DRAM} = f \times BlockSize / Bandwidth^{DRAM}$, and $L^{DRAM}$ is smaller on the A100 than on the other GPUs.
Specifically, as calculated from \cref{tab:evaluation_environment}, the value of $f / \mathrm{Bandwidth}^{DRAM}$ for the A100 is approximately one-third that of the other GPUs.
Therefore, on the A100, the negative impact of $S^{DRAM}$ is reduced.
At the same time, DPOB benefits from fewer kernel launches and improved latency hiding due to a larger number of concurrently active warps, leading to the best performance across many settings.
This interpretation is consistent with \cref{fig:stall_breakdown}, which presents the breakdown of stall cycles for selected kernels under the CKKS parameter setting $(4, 2^{16}, 30)$.
Although other GPUs exhibit a high fraction of \texttt{long} stalls (including $S^{DRAM}$) in DSOB, the A100 shows a smaller fraction, suggesting that DRAM access contention has a less pronounced impact on the A100.

Finally, we discuss why DPOB tends to achieve the best performance on the RTX 2080 Ti.
Because its L2 cache capacity is small, performance differences caused by memory-access behavior are less pronounced across the four strategies.
As shown in \cref{fig:memory_usage}, the L2 cache of the RTX 2080 Ti can fully accommodate DSOC and only some configurations of DPOC/DSOB.
In practice, as shown in \cref{fig:l2_cache_hit}, the RTX 2080 Ti exhibits lower L2 hit rates than the other GPUs, and the differences among the four strategies are also small.
Consequently, similar to the A100, DPOB tends to perform best across many settings due to lower kernel-launch overhead and better utilization of parallelism.

\subsection{Memory-related Profile}
\cref{fig:cache_hit} illustrates the distribution of L1 and L2 cache hit rates.
For the L1 cache, no significant differences are observed among the four strategies.
This is because the L1 cache capacity is small relative to the ciphertext size and is therefore insensitive to differences in inter-kernel dataflow optimization.

In contrast, for the L2 cache, the hit rates generally follow the order DSOC $>$ DSOB, DPOC $>$ DPOB, except on the RTX 2080 Ti.
This trend is consistent with the relative memory footprints of the approaches and aligns with the execution-time trends observed on the RTX 6000 Ada and RTX 4090.
On the RTX 2080 Ti, however, differences in L2 hit rate are minor because even the smallest-footprint strategy (DSOC) exceeds the available L2 cache capacity.

\subsection{Selection of $chunks$ in OutputChunked}
Finally, we evaluate the selection of $chunks$ in OC.
\cref{fig:best_chunks} shows the distribution of selected $chunks$ values in OC.
In OC, the number of output partitions ($chunks$) is chosen from 2 to 10 to minimize execution time.
As shown in \cref{fig:best_chunks_dsoc}, DSOC most frequently achieves optimal performance with $chunks = 2$ across many GPU and CKKS parameter combinations.
For DPOC, as shown in \cref{fig:best_chunks_dpoc}, larger CKKS parameters on the RTX 6000 Ada and RTX 4090 tend to favor $chunks = 4–6$ for optimal performance.

%% file: text/conclusion.tex
\section{Opportunities for Further Acceleration}

Based on the above performance analysis, we identify two promising directions for further improving GPU acceleration of CKKS.

The first direction, which is also the main focus of this study, is CKKS parameter-aware optimization.
Existing work typically adopts a uniform optimization strategy regardless of differences in CKKS parameter settings.
However, our results suggest that further performance improvements can be achieved by selecting the optimal dataflow strategy according to both the GPU architecture and the CKKS parameters.
In particular, the parameter $L$ changes depending on the number of executed multiplications even within a single workload.
This suggests that optimization strategies can be dynamically switched in response to changes in $L$ during execution.

The second direction is to explore DPOC-based optimization.
Although DPOC has not been adopted in prior work, it achieves the best execution time under certain CKKS parameter settings on the RTX 6000 Ada and RTX 4090.
Designing optimization strategies that incorporate DPOC may therefore lead to further performance improvements.

\section{Conclusion}
In this study, we demonstrated that in GPU acceleration of CKKS, the appropriate dataflow optimization strategy depends on the CKKS parameter configuration.
We classified dataflow optimization techniques proposed in existing work and conducted both qualitative and quantitative performance analyses of these approaches.
Our analysis revealed that the optimal strategy can vary depending on the CKKS parameters, and that the performance gap between different strategies can reach up to 1.98$\times$.
In addition, we found that the criteria for selecting an appropriate optimization strategy differ across GPU architectures.
Based on these findings, future work will work on accelerating whole CKKS workloads with parameter-aware and GPU architecture-aware optimization.